\documentclass[a4paper,11pt]{article}
\usepackage{pos}
\usepackage{amsmath}
\usepackage{graphicx}
\usepackage{mathrsfs}
\usepackage{multirow}
\usepackage{framed}

\def\beq{\begin{equation}}
\def\eeq{\end{equation}}
\def\bsp#1\esp{\begin{split}#1\end{split}}
\def\bal#1\eal{\begin{align}#1\end{align}}
\newcommand{\rd}       {{\mathrm{d}}}

\newcommand\tsigk[4]    {\sigma^{{\rm #1}^{#2} {\rm #3}^{#4}}}
\newcommand\dsig[1]    {\rd\sigma^{{\rm #1}}}
\newcommand\dsigk[4]    {\rd\sigma^{{\rm #1}_{#2} {\rm #3}_{#4}}}

\newcommand\dsiga[2]   {\rd\sigma^{{\rm #1,A}_{\scriptscriptstyle #2}}}

\title{{\tt NNLOCAL}: Fully Local Subtractions for Precision Predictions in Hadron Collisions }
\ShortTitle{{\tt NNLOCAL}}

\author*[a]{Pooja Mukherjee}
\author[b]{Vittorio Del Duca}
\author[c]{Claude Duhr}
\author[a,d]{Levente Fek\'esh\'azy}
\author[e]{Flavio Guadagni}
\author[f]{G\'abor Somogyi}
\author[g]{Francesco Tramontano}
\author[f]{Sam Van Thurenhout}

\affiliation[a]{II. Institut f\"ur Theoretische Physik, Universit\"at Hamburg, \\Luruper Chaussee
149, 22761, Hamburg, Germany}
\affiliation[b]{INFN, Laboratori Nazionali di Frascati,
00044 Frascati (RM), Italy}

\affiliation[c]{Bethe Center for Theoretical Physics, Universit\"at Bonn, \\
D-53115, Germany}
\affiliation[d]{Institute for Theoretical Physics, ELTE E\"otv\"os Lor\'and University, P\'azm\'any P\'eter s\'et\'any,\\
1/A, 1117, Budapest, Hungary}
\affiliation[e]{Physik-Institut, Universit\"at Z\"urich,\\ 8057 Z\"urich, Switzerland}
\affiliation[f]{HUN-REN Wigner Research Centre for Physics, \\Konkoly-Thege Mikl\'os u. 29-33, 1121 Budapest, Hungary}
\affiliation[g]{Dipartimento di Fisica Ettore Pancini, Universit\`a di Napoli Federico II and INFN - Sezione di
Napoli, Complesso Universitario di Monte Sant’Angelo Ed. 6, Via Cintia, 80126 Napoli, Italy}

\emailAdd{pooja.mukherjee@desy.de}
\emailAdd{Vittorio.DelDuca@lnf.infn.it}
\emailAdd{cduhr@uni-bonn.de}
\emailAdd{levente.fekeshazy@desy.de}
\emailAdd{fguada@physik.uzh.ch}
\emailAdd{somogyi.gabor@wigner.hun-ren.hu}
\emailAdd{francesco.tramontano@unina.it}
\emailAdd{sam.van.thurenhout@wigner.hun-ren.hu}

\abstract{This work extends the CoLoRFulNNLO subtraction method to address soft and collinear divergences in the computation of higher-order corrections for hadronic collisions. By utilizing universal local counterterms which can be integrated analytically over the unresolved phase space, we achieve numerically stable, fully-differential predictions. Our publicly available {\tt{NNLOCAL}} code serves as a proof-of-concept implementation, validated by calculating the NNLO cross-section for Higgs boson production in gluon-gluon fusion with no light quarks. }

\FullConference{17th International Symposium on Radiative Corrections: Applications of Quantum Field Theory to Phenomenology (RADCOR2025)\\
5-10 October 2025\\
Puri, India\\}

\begin{document}
\maketitle

\section{Introduction}
The Standard Model (SM) of particle physics remains a cornerstone in our understanding of elementary particles and their interactions. Over the past few decades, experimental validations, particularly at the Large Hadron Collider (LHC), have reaffirmed the robustness of the SM, notably through the groundbreaking discovery of the Higgs boson in 2012. However, as compelling as these findings are, they underscore the limitations of the SM in addressing critical phenomena such as dark matter, neutrino masses, and matter-antimatter asymmetries.

To navigate these challenges, precision tests of the SM have emerged as essential, especially in the realm of Quantum Chromodynamics (QCD), which plays a pivotal role in high-energy collisions. Theoretical predictions in collider physics rely heavily on perturbation theory, necessitating the incorporation of higher-order corrections to enhance accuracy. These corrections involve complex Feynman diagrams featuring additional real and virtual emissions, inherently leading to infrared (IR) divergences that must be carefully managed. A promising avenue for addressing these divergences is the use of local subtraction methods. By introducing counterterms that mimic the singular behavior of matrix elements in IR limits point-wise, these methods not only allow for the cancellation of divergences, but are also expected to lead to smoother numerical integrations. While established techniques exist at Next-to-Leading Order (NLO)~\cite{Frixione:1995ms,Catani:1996vz}, the extension of these methods to Next-to-Next-to-Leading Order (NNLO) remains an active area of research~\cite{Catani:1999ss,Gehrmann-DeRidder:2005btv,Caola:2017dug,Magnea:2018hab,Czakon:2014oma,Herzog:2018ily,Anastasiou:2022eym}.

In this work, we present an overview of the extension of the CoLoRFulNNLO method~\cite{Somogyi:2006db,Somogyi:2006da,DelDuca:2015zqa,DelDuca:2016csb,Somogyi:2020mmk,DelDuca:2024ovc,Guadagni:2026bby}, specifically its application to hadronic collisions. We detail the construction of counterterms based on the universal characteristics of QCD amplitudes in IR limits and their analytical integration over radiation phase space. Additionally, we introduce the {\tt NNLOCAL} code, a proof-of-concept implementation of this method, and showcase its capacity to compute NNLO corrections to Higgs boson production cross sections within the Higgs effective field theory (HEFT) framework. The following sections will provide a structured review of our methodology, results, and implications for future research in higher-order perturbative computations.

\section{Formalism of the Local Subtraction Method}

At fixed order in perturbation theory, a partonic cross section can be organised according to the number of real emissions and virtual loops. For a process $a + b \rightarrow X+k+m$ jets, with $X$ being the system of color-singlet particles and $k$ being the number of extra emissions compared to the Born-level interaction, the cross section is written as a sum of contributions with $k-l$ real emissions and $l$ virtual corrections. At NNLO, for which $k=2$, this naturally leads to the familiar decomposition into double-real, real-virtual, and double-virtual terms:

\beq
\tsigk{NN}{}{LO}{}_{ab} = 
    \int_{X+2+m} \underbrace{\dsigk{RR}{}{}{}_{ab} J_{X+m+2}}_{R_2V_0}
    +\int_{X+1+m}\underbrace{\left(\dsigk{R}{}{V}{}_{ab} + \dsigk{C}{1}{}{}_{ab}\right) J_{X+1+m}}_{R_{1}V_1}
    +\int_{X+m}\underbrace{\left(\dsigk{}{}{VV}{}_{ab} + \dsigk{C}{2}{}{}_{ab}\right) J_{X+m}}_{R_0V_2}\,.
\label{eq:tsigNkLO}
\eeq
Here, $\tsigk{NN}{}{LO}{}_{ab}$ denotes the NNLO partonic cross section for incoming partons $a,b$, evaluated for an infrared-safe observable defined by the measurement function $J$.  The first term represents the double-real contribution $\dsigk{RR}{}{}{}_{ab}$, describing tree-level configurations with two additional emitted partons and integrated over the 
$(X+2+m)$-particle phase space. The second term contains the real–virtual contribution 
$\dsigk{R}{}{V}{}_{ab}$, corresponding to one real emission at one loop, supplemented by $\dsigk{C}{1}{}{}_{ab}$ the collinear remnant , and integrated over the  $(X+1+m)$-particle phase space. Finally, the double-virtual contribution $\dsigk{}{}{VV}{}_{ab}$, describing two-loop corrections with Born-level kinematics, is combined with the remnant $\dsigk{C}{2}{}{}_{ab}$ and integrated over the $X+m$-particle phase space. These collinear remnants account for initial-state collinear singularities and ensure the cancellation of infrared divergences between real and virtual contributions. 

The first step of the CoLoRFulNNLO scheme is the local subtraction of infrared singularities from real radiation. We start from the $k$-fold real emission contribution 
$R_kV_0$, which develops divergences whenever one or more emitted partons become unresolved. To remove these singularities locally in phase space, we introduce subtraction terms $\mathcal{A}_k$ that reproduce the exact behaviour of the matrix element in the corresponding unresolved limits. In particular, the operator captures the limit in which all $k$ emitted partons become unresolved, while operators with lower indices account for configurations with fewer unresolved partons. By subtracting these approximations in a hierarchical manner, all infrared limits of the real-emission contribution are removed without double counting. This procedure can be written as,
\beq
(1-\mathcal{A}_1)(1-\mathcal{A}_2)\cdots (1-\mathcal{A}_k)[R_kV_0] = \prod_{i=1}^{k}(1-\mathcal{A}_i)[R_kV_0]\,,
\eeq
which is free of infrared singularities in all unresolved limits.

The subtraction terms introduced in the first step isolate the behaviour of the real-emission contribution in specific unresolved limits. Now, as a second step, any term involving the operator $\mathcal{A}_j$, which isolates a $j$-fold unresolved configuration, is integrated over the phase space of the unresolved partons. After integration, the resulting contribution has the kinematics of a configuration with $k-j$ real emissions and therefore must be combined with the corresponding $[R_{k-j}V_j]$ contribution. In this way, the infrared poles generated by the integration of the subtraction terms cancel analytically against the poles of the virtual corrections at the same underlying multiplicity. Then the rearrangement looks like the following:
\beq
\tsigk{N^k}{}{LO}{}_{ab,\text{Reg}} = \prod_{i=1}^{k}(1-\mathcal{A}_i)[R_kV_0] + \sum_{j=1}^{k}\bigg\{ R_{k-j}V_{j} - \int\prod_{i=1}^{j-1}(1-\mathcal{A}_i)(-\mathcal{A}_j)[R_kV_0]\bigg\}\,,
\eeq
Hence, iterating the first two steps--local subtraction of unresolved limits and the recombination of their integrated counterterms until all infrared singularities are removed--yields a fully infrared-finite cross section after the $k$-th iteration. This constitutes the general idea of the CoLoRFulNNLO scheme. Applying this framework at NNLO to color-singlet production ($m=0$), the cross section can be written as,

\beq
\bsp
&
\tsigk{NN}{}{LO}{}_{ab,\text{Reg}}  = 
    \int_{X+2}\left[
    \dsig{RR}_{ab} J_{X+2} - \dsiga{RR}{1}_{ab} J_{X+1} - \dsiga{RR}{2}_{ab} J_{X} + \dsiga{RR}{12}_{ab} J_{X} 
    \right]
\\ &\quad+
    \int_{X+1}\left\{
    \left[\dsig{RV}_{ab} + \dsigk{C}{1}{}{}_{ab} + \int_1 \dsiga{RR}{1}_{ab}\right] J_{X+1}
    -\left[\dsiga{RV}{1}_{ab} + \dsigk{C}{1,}{A}{1}_{ab} + \left(\int_1 \dsiga{RR}{1}_{ab}\right)^{\!{\rm A}_1}\right] J_{X}
    \right\}
\\ &\quad+
    \int_{X} \left\{
    \dsig{VV}_{ab} + \dsigk{C}{2}{}{}_{ab} + \int_2\left[\dsiga{RR}{2}_{ab} - \dsiga{RR}{12}_{ab}\right] + \int_1\left[\dsiga{RV}{1}_{ab} + \dsigk{C}{1,}{A}{1}_{ab}\right] + \int_1\left(\int_1 \dsiga{RR}{1}_{ab}\right)^{\!{\rm A}_1}
    \right\} J_{X}\,,
\esp
\label{eq:tsigNNLO}
\eeq
and the various approximate cross sections have the following interpretation:
\begin{itemize}
\item $\dsiga{RR}{1}_{ab}$ approximates the double real emission cross section $\dsig{RR}_{ab}$ in all single unresolved limits.
\item $\dsiga{RR}{2}_{ab}$ approximates the double real emission cross section $\dsig{RR}_{ab}$ in all double unresolved limits.
\item $\dsiga{RR}{12}_{ab}$ approximates $\dsiga{RR}{2}_{ab}$ in all single unresolved limits {\em and} $\dsiga{RR}{1}_{ab}$ in all double unresolved limits.
\item $\dsiga{RV}{1}_{ab}$ approximates the real-virtual cross section $\dsig{RV}_{ab}$ in all single unresolved limits.
\item $\dsigk{C}{1,}{A}{1}_{ab}$ approximates the collinear remnant $\dsigk{C}{1}{}{}_{ab}$ in all single unresolved limits.
\item $\left(\int_1 \dsiga{RR}{1}_{ab}\right)^{\!{\rm A}_1}$ approximates the integrated single unresolved approximate cross section $\int_1 \dsiga{RR}{1}_{ab}$ in all single unresolved limits.
\end{itemize}
With these subtractions, all three contributions on the right hand side of \eqref{eq:tsigNNLO} are rendered finite in four dimensions and can be computed numerically.
\section{Computation overview}

Having established the general structure of the NNLO partonic cross section in \eqref{eq:tsigNNLO}, we now turn to the construction and computation of these subtraction terms. The guiding principle for the construction of the counterterms is the universal factorization of QCD matrix elements in unresolved limits. 
\begin{align}
    \textbf{U}_{j}|M_{ab,m+j}(\{p\}_{m+j})|^2_{l-\text{loop}} \propto\sum_{i=0}^{l}\text{Sing}_{j}^{(i)}\times\underbrace{|M_{\hat{a}\hat{b},m}(\{\hat{p}\}_{m})|^2_{(l-i)-\text{loop}}}_{j ~\text{partons removed}}
\end{align}
These factorization properties determine the singular behaviour of real-emission contributions and provide the building blocks for defining subtraction terms that locally reproduce the infrared structure of the exact matrix elements. At the same time, care must be taken to avoid double counting in regions where several unresolved limits overlap.
\\
\\
At NNLO, such overlapping configurations are unavoidable, as multiple partons can become unresolved simultaneously, giving rise to soft–soft, collinear–collinear, and mixed soft–collinear limits. A consistent subtraction scheme must therefore correctly disentangle these overlapping singular regions, ensuring that each infrared configuration is subtracted exactly once while maintaining locality in phase space.
The detailed construction of the subtraction terms, together with the systematic treatment of overlapping unresolved limits, is presented in \cite{DelDuca:2025yph}.

\textcolor{black}{With the counterterms defined, we integrate them over the unresolved phase space, thereby ensuring the explicit cancellation of infrared poles among the double-real, real–virtual, and double-virtual contributions.} In the CoLoRFulNNLO framework, these integrated counterterms are constructed fully locally and evaluated analytically, allowing all singular limits and their overlaps to be handled in a systematic way.

Different integration strategies are employed depending on the structure of the subtraction term. For example, for the integration of the $\text{A}_1$ and $\text{A}_{12}$ counterterms, a direct integration approach is adopted, in which the phase space integrals are mapped to  parametric integrals and then evaluated to \textcolor{black}{\textit{multiple polylogarithms} (MPLs) \cite{goncharov2001multiplepolylogarithmsmixedtate} using {\tt PolyLogTools}\cite{Duhr:2019tlz}}. Due to the complexity of the integrands, particular care is required at intermediate stages of the calculation. \textcolor{black}{In particular, the non-trivial denominator structures are first linearized using partial fractioning; however, publicly available tools such as {\tt Apart} in Mathematica lead to long runtimes, rendering this step impractical. We overcome this limitation by developing a new univariate partial-fractioning routine, {\tt LinApart} \cite{Chargeishvili:2024nut}, which makes the procedure feasible}. The resulting expressions are then amenable to sector decomposition, and the singular behaviour is extracted and subtracted using the definition of plus distributions: 
\begin{equation}
(1-x_a)^{-1+m\varepsilon}= \frac{1}{m\varepsilon}\,\delta(1-x_a)
+\left[\frac{1}{1-x_a}\right]_+
+ m\varepsilon \left[\frac{\ln(1-x_a)}{1-x_a}\right]_+
+ \mathcal{O}(\varepsilon^2)\,.
\end{equation}
The details of these computations are presented in \cite{VanThurenhout:2024hmd,Fekeshazy:2025ktp}. For the computation of the $\text{A}_{2}$ counterterms, both direct integration methods and the method of differential equations are employed. In the latter approach, integration-by-parts identities are derived using cut propagators, supplemented by additional propagators in order to complete the basis of integral families. Once this basis is established, the \textit{Integration-by-parts} (IBP) reduction yields a set of master integrals, for which systems of differential equations are constructed and subsequently solved. \textcolor{black}{The details of these calculations will be discussed in a future publication}.

A key outcome of this approach is that all integrated subtraction terms are obtained in fully analytic form. After simplification, the results can be expressed in terms of classical polylogarithms, including algebraic letters with square roots appearing in the finite parts. The cancellation of infrared poles from the integrated counterterms and the virtual amplitudes is verified analytically. 
\section{Numerical Implementation}
The full subtraction formalism described above, including all analytically integrated counterterms, has been implemented in the parton-level Monte Carlo program \texttt{NNLOCAL}~\cite{DelDuca:2024ovc}. The code provides a fully local and analytic NNLO subtraction framework and is publicly available at
\url{https://github.com/nnlocal/nnlocal}
. In addition to the numerical evaluation of infrared- and collinear-safe observables, \texttt{NNLOCAL} includes dedicated tools to explicitly monitor the cancellation of kinematic singularities and virtual poles, as well as facilities for efficient phase-space generation and Monte Carlo integration.

In its current public version, the code is configured for Higgs boson production in hadron collisions within the Higgs effective field theory (HEFT) approximation, considering the gluon-fusion channel with no light quarks ($n_f$=0). The implementation has been validated by computing the inclusive NNLO cross section and performing tuned comparisons with the independent code \texttt{n3loxs}~\cite{Baglio:2022wzu}, after excluding quark-induced channels. We find perfect agreement between the two calculations, providing a stringent check of the subtraction framework and its numerical realization.

Since \texttt{NNLOCAL} is fully differential in all final-state momenta, it can also be used to compute differential observables through user-defined analysis routines. As an illustration, we present the rapidity distribution of a Higgs boson with mass $m_H$=125 GeV at the 13~TeV LHC in the HEFT approximation with $n_f$=0. The results demonstrate very good numerical convergence and stability across the rapidity range. For moderate bin sizes, the Monte Carlo uncertainties remain under good control, while finer binning leads to visible fluctuations associated with misbinning effects, which can be mitigated using robust estimators for the bin values. For a detailed discussion of misbinning effects, please refer to \cite{DelDuca:2024ovc}.
\begin{figure}[ht] 
  \centering
  \includegraphics[width=1.0\textwidth]{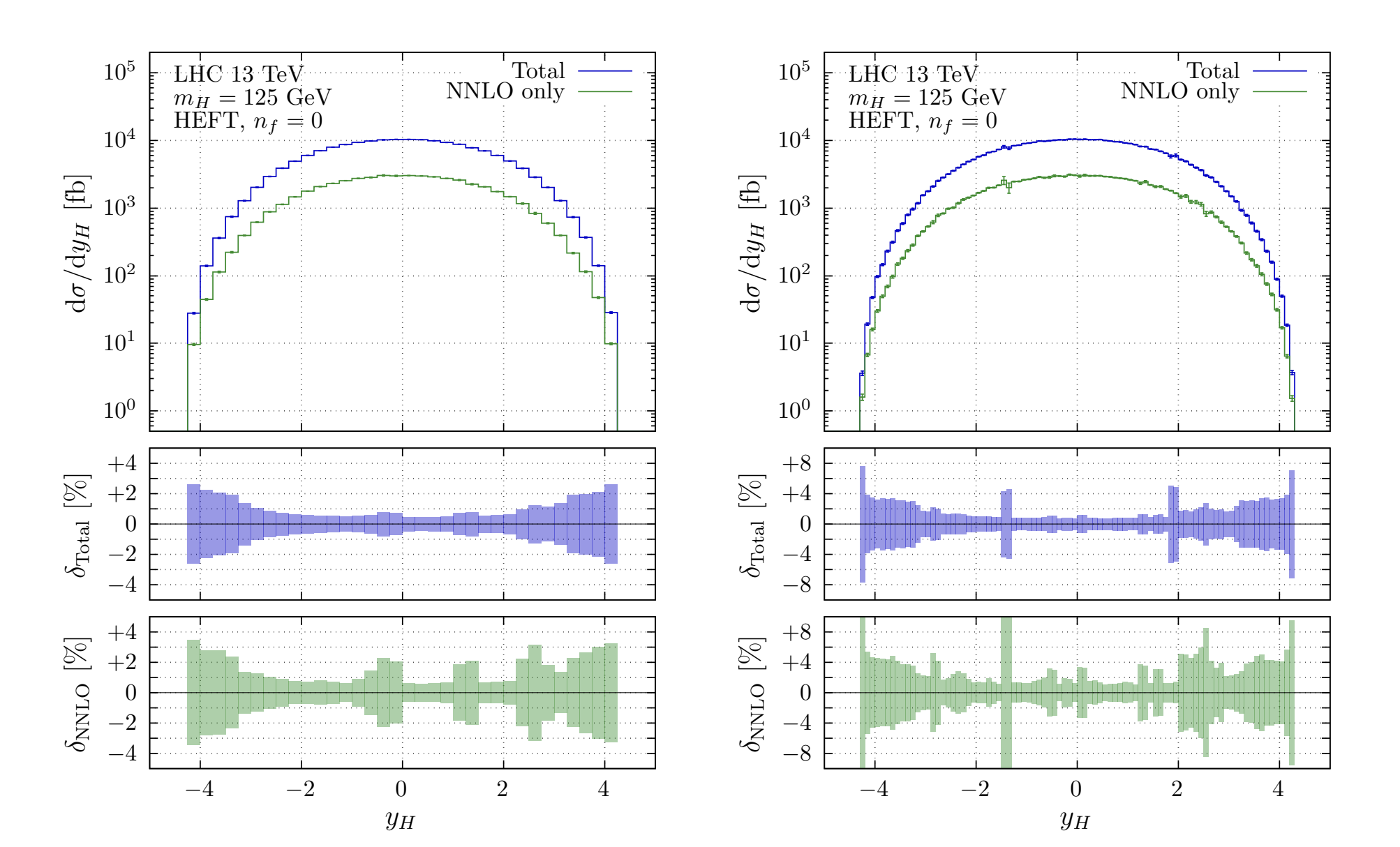} 
  \caption{The rapidity distribution of the Higgs boson at NNLO in the HEFT with $n_f=0$ light quarks. The distribution on the left has a bin width $\Delta y=0.25$, while the one on the right has a bin width of $\Delta y=0.1$. The lower panels show the relative error for the total distribution and the NNLO contribution. The error bands shown are the estimated Monte Carlo uncertainties. Figures taken from \cite{DelDuca:2024ovc}.}
  \label{fig: rapidity distribution}
\end{figure}
The reported results were obtained on a MacBook Pro laptop equipped with an M2 processor and 8 CPU cores, with a typical runtime of about one hour. Ongoing developments of the code include the implementation of all partonic channels for Higgs boson production and the extension to additional processes, such as vector boson production in hadron collisions. In particular, the full set of double-real emission subprocesses for $pp\rightarrow H$ has already been implemented, allowing for detailed studies of the regularized double-real contribution as an additional non-trivial validation of the framework.
\section{Conclusion}
In this contribution, we have reported on the extension of the CoLoRFulNNLO subtraction method to the production of color-singlet final states in hadronic collisions. The infrared singularities arising at NNLO are regularized by local subtraction terms constructed through a careful extension of the known QCD infrared factorization formulae over the full real-emission phase space. This allows for a fully local treatment of all unresolved configurations over the phase space.

By applying a combination of complementary integration techniques, we have obtained fully analytic expressions for the integrated subtraction terms relevant to color-singlet production at NNLO. This analytic control allows for explicit checks of the cancellation of virtual poles and facilitates an efficient numerical implementation.

The subtraction formalism has been implemented in the publicly available parton-level Monte Carlo program \texttt{NNLOCAL}. While the current public version is a proof-of-concept with a limited scope, ongoing developments aim at extending the code to include all partonic channels and additional processes of phenomenological interest. 
\section*{Acknowledgments}
This work has been supported by grant K143451 of the National Research, Development and Innovation Fund in Hungary and the Bolyai Fellowship program of the Hungarian Academy of Sciences. The work of C.D.\ was funded by the European Union (ERC Consolidator Grant LoCoMotive 101043686). The work of P.M\ was supported by the ERC Advanced Grant 101095857 Conformal-EIC. Views and opinions expressed are however those of the author(s) only and do not necessarily reflect those of the European Union or the European Research Council. Neither the European Union nor the granting authority can be held responsible for them. The work of L.F.\ was supported by the German Academic Exchange Service (DAAD) through its Bi-Nationally Supervised Scholarship program.
\bibliographystyle{unsrt}
\bibliography{biblio}
\end{document}